# Tuning the Reactivity of Nanoenergetic Gas Generators Based on Bismuth and Iodine oxidizers

Mkhitar A. Hobosyan and Karen S. Martirosyan*

Department of Physics and Astronomy, University of Texas at Rio Grande Valley, Brownsville, TX 78520, USA

*karen.martirosyan@utrgv.edu

**Abstract**

There is a growing interest on novel energetic materials called Nanoenergetic Gas-Generators (NGGs) which are potential alternatives to traditional energetic materials including pyrotechnics, propellants, primers and solid rocket fuels. NGGs are formulations that utilize metal powders as a fuel and oxides or hydroxides as oxidizers that can rapidly release large amount of heat and gaseous products to generate shock waves. The heat and pressure discharge, impact sensitivity, long term stability and other critical properties depend on the particle size and shape, as well as assembling procedure and intermixing degree between the components. The extremely high energy density and the ability to tune the dynamic properties of the energetic system makes NGGs ideal candidates to dilute or replace traditional energetic materials for emerging applications. In terms of energy density, performance and controllability of dynamic properties, the energetic materials based on bismuth and iodine compounds are exceptional among the NGGs. The thermodynamic calculations and experimental study confirm that NGGs based on iodine and bismuth compounds mixed with aluminum nanoparticles are the most powerful formulations to date and can be used potentially in microthrusters technology with high thrust-to-weight ratio with controlled combustion and exhaust velocity for space applications. The resulting nano thermites generated significant value of pressure discharge up to 14.8 kPa m$^3$/g. They can also be integrated with carbon nanotubes to form laminar composite yarns with high power actuation of up to 4700 W/kg, or be used in other emerging applications such as biocidal agents to effectively destroy harmful bacteria in seconds, with 22 mg/m$^2$ minimal content over infected area.

1. Introduction

Nanoenergetic Gas Generators (NGGs) are thermite based nano-structured formulations, where the gas generation ability of the system is most important characteristic, and is measured by the ability of the system to generate pressure discharge during very short time (order of microseconds). The main difference of nano-structured thermites from so-called traditional thermites is the reduced particle size, where at least one component of the thermite should be in nano-size particle domain of less than 100 nm [1-3]. Thermites are well known pyrotechnic composite mixtures of metal powder and metal or non-metal oxide, which can produce an exothermic reaction also known as aluminothermic reaction. The traditional thermites have been known since 19[th] century [4].

The class of compositions in thermites can be very diverse, where the metal powders act as a fuel, and the oxidizers are various oxides such as traditional metal oxides including $Fe_2O_3$, $Co_3O_4$, $MoO_3$, etc [5-7]. Other highly powerful oxidizers include Nanoenergetic Gas Generators [2, 8] constituents such as $I_2O_5$ [9-12] and $Bi_2O_3$ [13-16]. The metal fuel in thermites is usually aluminum powder, although occasionally magnesium is also utilized as a fuel metal [17]. The reason behind the popularity of Al as a fuel is that it is one of the cheapest metals to produce, it is highly energetic and reactive as well as exhibits relatively low melting temperature (660 $^o$C). This implies that the metal can melt and react with oxidizers in liquid form, tremendously accelerating the reaction rate in comparison to solid-solid reactions. Although aluminum is highly reactive, the aluminum oxide layer that naturally forms on aluminum particles, protects the aluminum core and makes it safer to use even 20-100 nm particle size.

The thermites at nano-scale have very high chemical reaction rate in comparison to thermites at micro-scale. When reducing the particle size to nano-meter domain, the reaction propagation velocity can increase by up to 2-3 orders of magnitude for some systems [7, 18-20], making them comparable or even better than the traditional energetic materials such as pentaerythritol tetranitrate (PETN), 2,4,6-trinitrotoluene (TNT), or 1,3,5-trinitroperhydro-1,3,5-triazine (RDX). The reason behind is that the nano-thermite systems have higher volumetric energy density compared to traditional energetic materials [21]. However, this large energy capacity can be only effectively utilized, if the thermites are at nano-scale. When reducing the thermite reactants particle size to nano-sized domain, the contact between reagents particles becomes much more intimate than in micron-sized particles, and, moreover, the size reduction decreases the diffusion and transport limitations [22-24]. Thus, the successful energy utilization in nano-thermite mixtures gives them great potential to become the next generation energetic materials both in military (propellants, primers, pyrotechnics) and civilian applications (energy, mining, etc) [25-27]. The NGGs can create shock waves with velocities up to 2500 m/s [13, 21], which is opening perspectives for more applications such as medicine, biological sciences, material processing, manufacturing, and microelectronic industries. We should note that the energy capacity has critical role in determining the pressure discharge capability of energetic materials as demonstrated by Martirosyan et al in [21]. Bismuth and iodine oxides mixtures with aluminum, have one of the highest energetic capacities per volume, among common nano-thermite systems. Both bismuth oxide and bismuth hydroxide have shown excellent performance in NGG formulations Al-$Bi_2O_3$ [13, 16] and Al-$Bi(OH)_3$ [28]. The corresponding maximum pressure x volume (PV) per mass (PV/m) value for Al-$Bi_2O_3$ nano-structured formulation was 8.6 kPa $m^3$/g [13, 16], while for the Al-$Bi(OH)_3$ formulation the corresponding value was 5.6 kPa $m^3$/g. Although the energetic capacity per volume or per mass for Al-$Bi(OH)_3$ system is lower than that for Al-$Bi_2O_3$ formulation, it can create twice more gaseous products per mass [28]. This could be advantageous for hydroxide-based systems in comparison to oxide-based thermites, especially in applications requiring large amounts of gaseous products to be released upon charge ignition, such as micro-thrusters and micro-propulsion platforms for space and terrestrial applications [29]. The iodine pentoxide based system has shown even higher PV/m value of 14.8 kPa m$^3$/g in the nano-structured formulation Al-I$_2$O$_5$, and can be used not only as propellant in microthrusters [30, 31], but also as a very efficient biocidal agent due to highly active atomic iodine released during nano-thermite reaction [32].

## 2. Thermodynamic considerations

For the estimation and control of the thermodynamic properties of the energetic systems and reaction products, knowledge of the adiabatic reaction temperature and equilibrium concentration of solid, liquid and gas phases is needed. The thermodynamic estimation of the equilibrium composition of multicomponent multiphase systems requires minimization of the thermodynamic free energy (G) subject to mass and energy balances [33]. "Thermo" software was used for thermodynamic calculations [34], which has thermochemical database for approximately 3000 compounds. In addition, we used the thermo-chemical software HSC Chemistry-7, which can predict equilibrium compositions and can calculate adiabatic temperature, if input and output species and their amounts are defined. HSC-Chemistry-7 has database of over 25000 compounds, based on which it can also calculate the theoretical heat balances, by taking molecular amounts of components using the reaction equations at system equilibrium.

The reaction adiabatic temperature and composition of the equilibrium products can be estimated by minimizing the thermodynamic potential. For a system with N(g) gas and N(s) solid number of components, at constant pressure, the concentrations of equilibrium phases can be expressed as:

$$F(\{n_k\},\{n_s\}) = \sum_{k=1}^{N^{(g)}} n_k \left(ln\frac{p_k}{p} + G_k\right) + \sum_{l=1}^{N^{(s)}} n_l G_l \qquad (1)$$

Where $p_k$ is the partial pressure of the k`th gas-phase component, while $n_l$ and $G_l$ are the number of moles and molar Gibbs free energy of components. The adiabatic combustion temperature, $T_c^{ad}$, is determined by total energy balance:

$$\sum_{i=1}^{N_0} H_i(T_0) = \sum_{k=1}^{N^{(g)}} n_k H_k(T_c^{ad}) + \sum_{l=1}^{N^{(s)}} n_l H_l(T_c^{ad}) \qquad (2)$$

Where the enthalpy of each component is

$$H_i(T) = \Delta H_{f,i}^0 + \int_{T_0}^{T} c_{p,i} dT + \sum \Delta H_{s,i} \qquad (3)$$

and $\Delta H_{f,i}^0$ is the heat of formation at 1 atm and reference temperature $T_0$, $c_{p,i}$ is the heat capacity, and $\Delta H_{s,i}$ is the heat of s`th phase transition for components [35]. The combination of both above mentioned software programs made it possible to estimate the equilibrium composition, adiabatic temperature, gas generation, and standard enthalpy of formation for each examined system.

The reason of exceptional performance for bismuth and iodine based NGGs can be that the boiling temperature of oxide or hydroxide metal (iodine or bismuth) is significantly lower than the combustion adiabatic temperature, which contributes to gaseous products during combustion and improves the pressure generation ability [2, 32, 36]. The adiabatic combustion temperature of common nano-thermite formulations at stoichiometric ratios along with oxidizer metal boiling temperature and normalized gas generation are presented in Table 1. It can be seen, that boiling temperature of final product for systems based on iodine and bismuth oxidizers, is significantly lower than adiabatic combustion temperature. They generate significant amounts of gaseous products, with the highest number for iodine pentoxide based NGGs. The other formulations have the oxidizer metal boiling point higher than the combustion adiabatic temperature, and the gas generation is due to partial decomposition of reaction product aluminum oxide into gaseous aluminum oxides such as $Al_2O(g)$. Therefore, lower boiling point of oxidizer metal appears to be one of the most important factors for high pressure discharge value.

**Table 1.** Boiling point of metal, maximum combustion adiabatic temperature, and normalized gas generation per initial thermite mass, for thermite formulations.

| Combustion system | Boiling point of metal, (K) [37] | Maximum combustion adiabatic temperature, (K) | Normalized Gas Generation, (L/g) |
|---|---|---|---|
| $10\ Al + 3\ I_2O_5$ | (I) 457 | 3827 | 3.1 |
| $2\ Al + Bi(OH)_3$ | (Bi) 1833 | 2965 | 2.1 |
| $2\ Al + Bi_2O_3$ | (Bi) 1833 | 3279 | 1.1 |
| $8Al + 3\ Co_3O_4$ | (Co) 3200 | 3174 | 0.6 |
| $2\ Al + MoO_3$ | (Mo) 4921 | 3808 | 0.4 |
| $2\ Al + Fe_2O_3$ | (Fe) 3134 | 3130 | 0.3 |

The NGGs perform better when both fuel and oxidizer are at nanostructured scale. The fuel Al can be purchased commercially at 100 nm average size (Sigma Aldrich). This size is not very pyrophoric, and is covered by approximately 4-5 nm aluminum oxide layer, which allows to safely mix with oxidizers.

The thermite reactions can be accelerated if the oxide layer is removed from Al nanoparticles. At high heating rates, which are characteristic to thermite types of reactions, the aluminum oxide can be removed with polytetrafluoroethylene (PTFE). It is interesting to note that the PTFE can react with the outer $Al_2O_3$ layer on aluminum particles, which is an exothermic process and can increase the oxidation reaction rate, and ultimately can improve the energy and gas discharge in PTFE containing thermites [38]. The PTFE activated thermites were successfully utilized for lunar regolith consolidation purposes [39, 40]. We should emphasize, that the activation of thermite reactions with PTFE is useful when the main system has lower combustion adiabatic temperature than the formulation Al-PTFE, such as the system $Al_2O_3$-

PTFE. Moreover, it should be noted that some components which can react with PTFE, may change the stoichiometric balance of Al-PTFE reaction and reduce the combustion temperature. Table 2 summarizes the results for the calculated adiabatic temperature for each system with or without PTFE. As can be seen, the addition of Al-PTFE system greatly increases the combustion adiabatic temperature for $Al_2O_3$-PTFE system from 1425 K to 3075 K. However, the addition of Al-PTFE formulation in systems containing bismuth oxide, bismuth hydroxide and iodine pentoxide has negative contribution, and reduces the combustion adiabatic temperature by several hundred degrees. The possible reason is due to the partial reaction between oxidizers ($Bi_2O_3$, $Bi(OH)_3$ and $I_2O_5$) with PTFE, which shifts the stoichiometric balance of Al-PTFE formulation, and reduces the combustion temperature. Thus, PTFE can be used to activate regolith-thermite reactions due to the fact that regolith contains significant amount of $Al_2O_3$. Thus, the ($Al_2O_3$-PTFE)-(Al-PTFE) system has higher energetic resources, as discussed above, and activation effect of Al-PTFE formulation is significant. However, PTFE should not be used in case of iodine and bismuth based oxidizers due to negative contribution in combustion temperature, as presented in Table 2.

**Table 2**. The adiabatic combustion temperature for thermites based on PTFE, iodine and bismuth compounds.

| Combustion system | Adiabatic temperature, K |
|---|---|
| PTFE-Al | 3587 |
| PTFE-$Al_2O_3$ | 1425 |
| (PTFE-$Al_2O_3$)-(PTFE-Al) | 3075 (large increase) |
| $Bi_2O_3$-Al | 3284 |
| ($Bi_2O_3$-Al)-(PTFE-Al) | 2552 (decrease) |
| $Bi(OH)_3$-Al | 2970 |
| ($Bi(OH)_3$-Al)-(PTFE-Al) | 2346 (decrease) |
| $I_2O_5$-Al | 3830 |
| ($I_2O_5$-Al)-(PTFE-Al) | 3423 (decrease) |

## 3. Preparation of iodine and bismuth based oxidizers at nano-scale

As mentioned above, the best performance for NGGs can be expected when not only fuel, but also oxidizers are at nano-scale. In addition to the particle size, the particle shape may also play important role when estimating the contact area between reagents, which is one of the most important factors for the reaction rate. The shape of Al fuel nano-particles is spherical, while the oxidizers can be prepared by various shape and size, which can have significant effect on NGG performance during pressure discharge characterizations [36]. Although $Bi_2O_3$, $Bi(OH)_3$ and $I_2O_5$ are commercially available as micro-meter sized powders, the nano-sized powder preparation can be challenging, especially for iodine pentoxide. The high energy ball milling allows to receive iodine pentoxide nano-rods [12] and bismuth hydroxide sub-micrometer and nano-sized particles [28], while bismuth oxide can be received at nano-scale using solution combustion method [13]. Bismuth oxide and hydroxide with various shape and size can be received using microfluidic synthesis approach [36].

For iodine pentoxide nano-particle preparation, the commercial iodine pentoxide micro sized particles with 98 % purity (Sigma Aldrich) can be used to produce $I_2O_5$ nano-rods structures. The particles compositions are very sensitive from environmental variation and, thus, sample storage-handling-preparation should be performed under nitrogen environment in glove-box, and iodine pentoxide nanoparticles should be stored and used under reduced humidity environment to protect formation of hydrated iodine pentoxide due to water molecules absorption by $I_2O_5$. For mechanical milling of micro-sized particles (Figure 1a), a High Energy Shimmy Ball Mill (HSF-3, MTI Co) was used. The ball to powder mass ratio in the milling container was 4:1. The particle size of iodine pentoxide was regulated by time controlling of high energy ball milling process via tuning the applied energy dose which is transferred to the milling media. The details of estimation of energy dose transferred to particles can be calculated using the method described in [12]. The total energy transferred per unit mass of powder, or the specific energy dose, $D_E$, [J/kg], is given by $D_E = \frac{NtE-(Q+q)}{m_p}$, where N is collision frequency [Hz]; t is the processing time [s]; $m_p$ is the mass of the powder batch [kg], Q is the thermal heat loss to milling container (stainless steel), calculated as $Q = c_{ps} M \, \Delta T$, where $c_{ps}$ is the heat capacity of stainless steel, M is the combined mass of metallic container and balls, ΔT is the change in temperature. q is the thermal heat loss to powder batch calculated as $q = c_{pp} m_p \, \Delta T$, where $c_{pp}$ is the heat capacity of powder being treated.

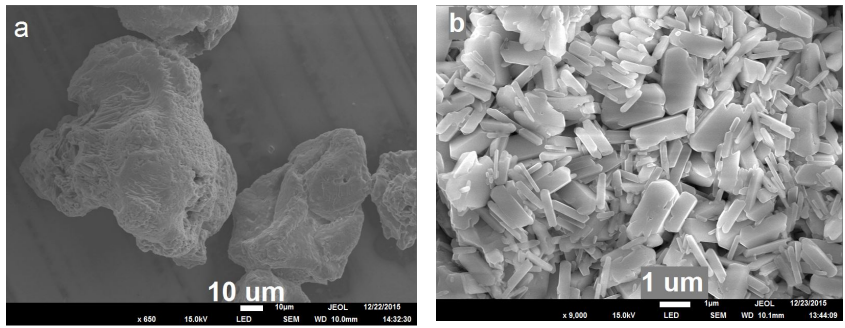

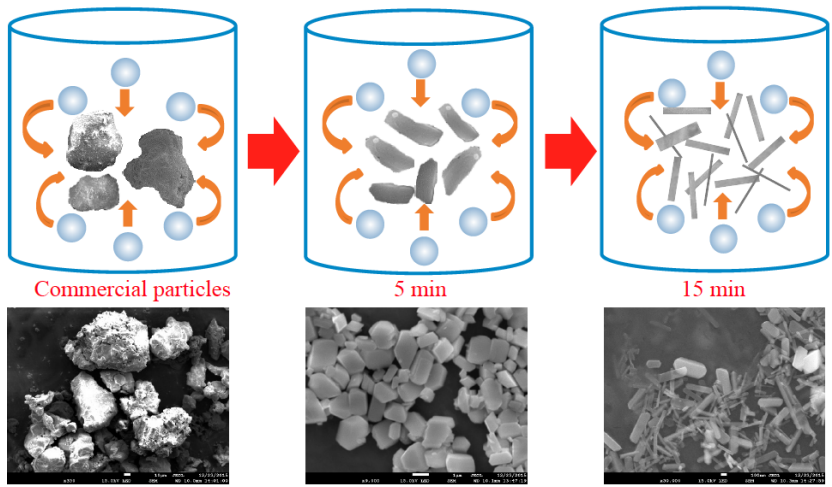

**Figure 1.** (a) Scanning Electron Microscopy (SEM) images of typical commercial iodine pentoxide particles; (b) Particles after 10 minutes of mechanical treatment; (c) schematic representation of mechanical treatment, showing the particle morphology after each step.

The $I_2O_5$ particles during mechanical treatment are gradually transforming into nano-rods (Figure 1 b). In this case, 15 minutes of mechanical treatment converted the iodine pentoxide micro-meter sized particles into nano-rods, shown in schematics in Figure 1c.

For the bismuth hydroxide, the mechanical treatment method is working similarly [28]. The initial micro-meter sized particles of commercially available (Acros Organics, 99 % purity) bismuth hydroxide (Figure 2a) are converted into sub-micrometer and nano-sized particles (Figure 2b).

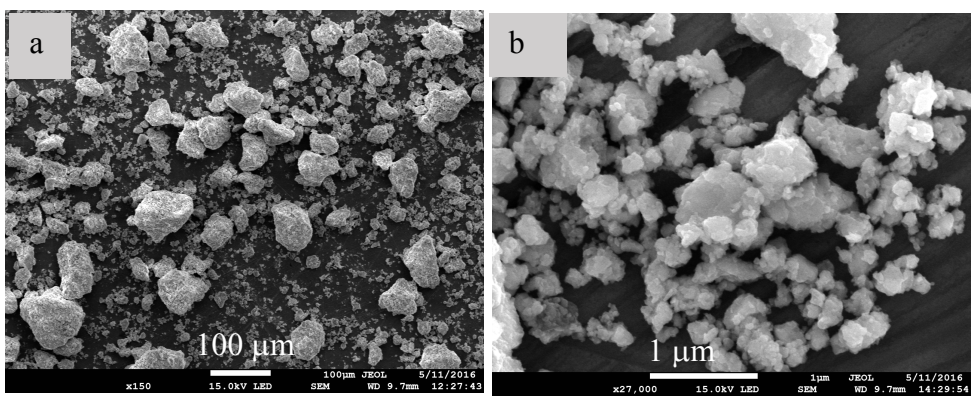

**Figure 2.** Commercially available bismuth hydroxide particles before (a) and after (b) mechanical treatment in high energy ball mill.

For bismuth oxide particle preparation, microfluidic synthesis approach (Figure 3a) was utilized to produce particles with various shape and estimate the dependence of pressure discharge of NGGs on oxidizer particle shape and size. $Bi_2O_3$ structures which resemble flowers, brushwood-like structures and bowties, were successfully synthesized using polyethylene glycol (PEG) with various molecular length as a reaction media [36]. The surfactant PEG molecular length was very important during formation of particle morphology, where PEG with 200

Molecular Weight (MW) results in production of 1-2 μm flower-like structures, which have self-assembled layers of petals with thickness of 20-30 nm (Figure 3b). Utilization of PEG with 8000 MW results in bowtie and brushwood-like structures, which are highly crystalline with up to 60 μm size (Figure 3 c, d). The nano-thermite mixtures prepared with flower-shaped particles as oxidizers, showed higher pressure discharge values than the mixtures prepared with bowtie- and brushwood like particles.

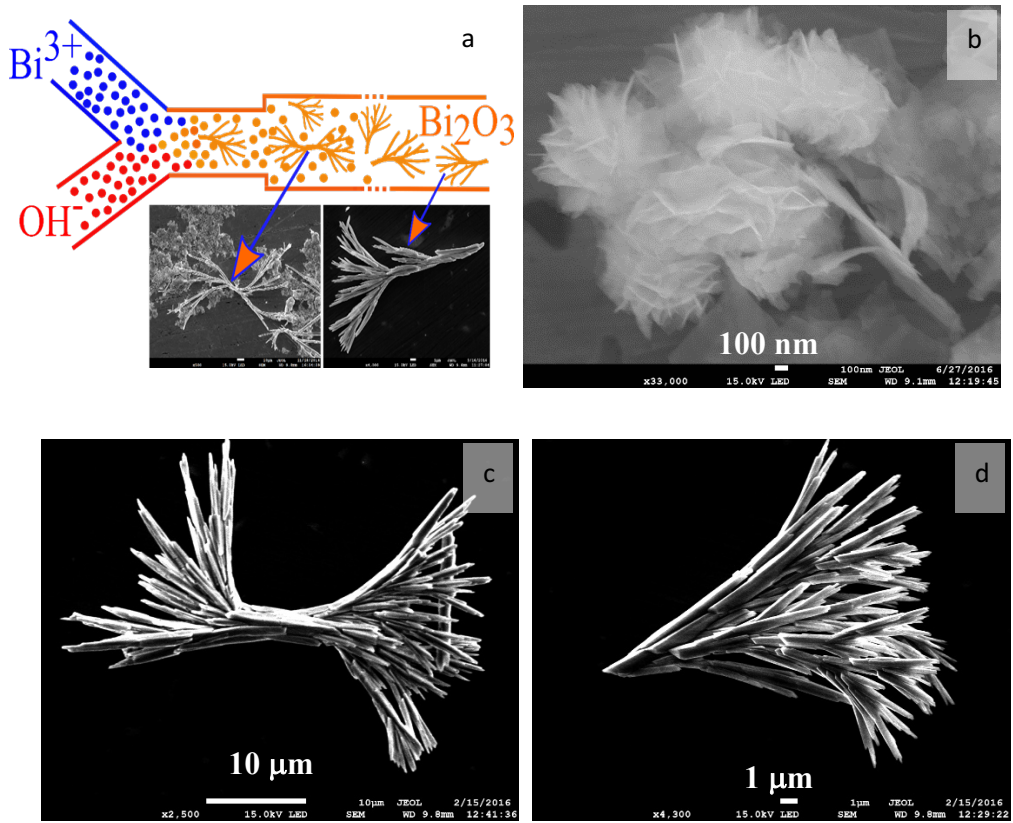

**Figure 3.** (a) Schematics of microfluidic synthesis, and SEM of $Bi_2O_3$ product utilizing (b) PEG-200 surfactant, resulting in flower-like particles, and (c, d) PEG-8000 surfactant, showing bow-tie like and brushwood-like structures.

In all cases, NGGs Al-$I_2O_5$, Al-$Bi_2O_3$ and Al-$Bi(OH)_3$ prepared with nano-sized oxidizer particles generated much higher value of pressure discharge, than the mixtures prepared with commercial micro-meter sized oxidizer. Table 3 is summarizing the maximum pressure discharge value for each composition.

**Table 3.** Maximum pressure generation for NGGs prepared with iodine and bismuth oxidizers with commercially available micrometer-sized, and nano-sized particles.

| Combustion system | Maximum pressure for commercial oxidizer, in 0.345 L volume, 0.2 g charge, (MPa/g) | Maximum pressure for nano-size oxidizer, in 0.345 L volume, 0.2 g charge, (MPa) | Normalized Pressure for nano-sized oxidizer, PV/m, (kPa*$m^3$/g) |
|---|---|---|---|
| Al - $I_2O_5$ | 4.4 | 8.68 | 14.8 |
| Al - $Bi(OH)_3$ | 3.33 | 4.34 | 5.6 |
| Al - $Bi_2O_3$ | 1.2 | 2.86 | 4.9 |

Thus, the NGGs based on iodine and bismuth oxidizers exhibited significant pressure discharge properties, which can be finely tuned not only by the size, but also by the shape of oxidizer particles. These tunable pressure generation abilities, as well as the generation of gaseous iodine in the case of Al-$I_2O_5$ formulation, were successfully utilized in the following emerging applications.

**4. Emerging applications of NGGs based on iodine and bismuth oxidizers**

As was outlined above, the rapid generation of copious amounts of gaseous products, which provides extremely high pressure in microseconds, places the NGGs based on iodine and bismuth components in a unique position. Aside from traditional applications as thermites in pyrotechnics, welding, metal cutting, elementary metal production, etc, they can be used in micropropulsion platforms as new types of propellants, as well as in hybrid actuators and biocidal applications in the case of Al-$I_2O_5$ system. Each of these new applications are detailed below.

*4.1 Nano-energetic micropropulsion systems*

The micropropulsion systems utilize Microelectromechanical System (MEMS) technology, which are integrated with high energy density nanoenergetic materials as solid propellants. These microthrusters can be used as single units or in arrays, in nano- and micro-satellites, for applications in payload delivery, navigation, guidance, stabilization, various maneuvering tasks, etc. [31, 41, 42]. Due to the complex structure and operation of microthrusters with liquid propellants [43, 44], the solid propellants are preferred, which have advantages such as simplicity, adequate functionality, safety, etc. The impulse control should be in mNs resolution [45]. The burning time of propellant in micro chamber is a few milliseconds, as the charge mass is in order of micrograms to milligrams. Since, the burning time of propellant in microthruster is very short, the polymeric materials can be preferred as structural materials to print 3D-optimized plastic microthrusters, which are lightweight and robust [29]. The thermal insulation in polymers is exceptionally good due to low thermal conduction, therefore nozzle and polymer chamber are not deformed and not melted. The same polymer microthrusters were tested multiple times, and, the impulse variations were within 10 %.

For the microthruster operation, the total impulse $I_{total}$ and specific impulse $I_{sp}$ are the key characteristics, and they are given as

$$I_{\text{total}} = \int_0^{t_c} F dt \qquad (4)$$

$$I_{sp} = \frac{I}{mg} \qquad (5)$$

where $t_c$ is combustion time; $m$ is the the propellant mass; $F$ is the thrust.

The specific impulse is proportional to combustion temperature ($T_c$) of gas products in the chamber, and inversely related to the average molecular mass of the gaseous products during combustion $W_g$. [46]

$$I_{sp} \sim \sqrt{\frac{T_c}{W_g}} \qquad (6)$$

Thus, in order to receive high specific impulse, the combustion temperature should be large, and the average molecular weight of combustion gaseous products should be low. In microthrusters the heat losses are significant, as the propellant mass is very small, which results in reduced specific impulse values compared to massive rocket engines.

It should be noted that the traditional energetic materials such as nitrocellulose, are not suitable to be used in microthrusters due to insufficient thrust generation at extremely small charge mass, which is a requirement for microthrusters. Most solid fuels in a large rocket engine are not suited for microthrusters due to ignition inconsistencies, encapsulation inadequacy, safety, etc. For example, the well-known hydroxyl-terminated polybutadiene (HTPB) - ammonium perchlorate (AP) is not sensitive sufficiently and cannot be used in microthrusters alone. A separate ignitor is necessary, which complicates design and operation [47]. The examined hybrid composite solid rocket fuels include either HTPB/AP compositions [45, 47, 48], or traditional energetic materials such as nitrocellulose (NC) combined with thermite types of formulations such as Al-$Bi_2O_3$-NC [49] and Al-CuO-10wt% NC [50]. The addition of NC improves the thrust generation ability of thermites until some small (2.5-10 wt. %) amount, after which the increase of NC has negative effect. We note, that the thermites can be also combined with HTPB or other binding polymers, which should generate desired thrust and tunability by ranging the concentration of thermites in polymers. The addition of thermite in HTPB should increase sensitivity and provide adequate thrust abilities, without separate ignitors.

As discussed above, the nanoenergetic compositions based on iodine and bismuth compounds, are very energetic and release large amounts of gaseous products. The reaction Al-$Bi_2O_3$ has larger heat release than Al-$Bi(OH)_3$, but the later produces almost twice the amounts of gaseous products than Al-$Bi_2O_3$ (Table 1). When comparing the theoretical specific impulse estimates by equation (6), the reaction Al-$Bi_2O_3$ yields $\sqrt{\frac{T_c}{W_g}} = 3.96$, while this number for Al-$Bi(OH)_3$, is $\sqrt{\frac{T_c}{W_g}} = 5.65$, which is even higher than the number for reaction Al-$I_2O_5$, calculated to be $\sqrt{\frac{T_c}{W_g}} = 5.5$. Thus, the bismuth hydroxide NGG formulation has the best propellant energetics. We have tested all 3 systems, and Al-$Bi(OH)_3$ provided the best thrust generation abilities and specific impulse [29-31].

In order to test the thrust generation abilities of Al-$Bi(OH)_3$ system, single microthrusters and thruster arrays were designed and printed, using thermoplastic polymer Acylonitrile Butadiene Styrene. Microthrusters were printed by a 3D printer "Printrbot Simple Metal" (Figure 4, a) with various nozzle geometries (outer-to-throat diameter ratios). The microthruster arrays can be printed with precise dimensions as presented in Figure 4, b. The evaluation of thrust generation for microthrusters, which were integrated with NGGs, is performed using the Phidget force sensors with bridgeboard as presented in Figure 4, c. The force sensor was calibrated before each measurement to ensure the measurement accuracy. To eliminate the effect of high-temperature igniter on sensing and imaging, the ignition of microthrusters was initiated using blue laser with power output of 1.4 W [29].

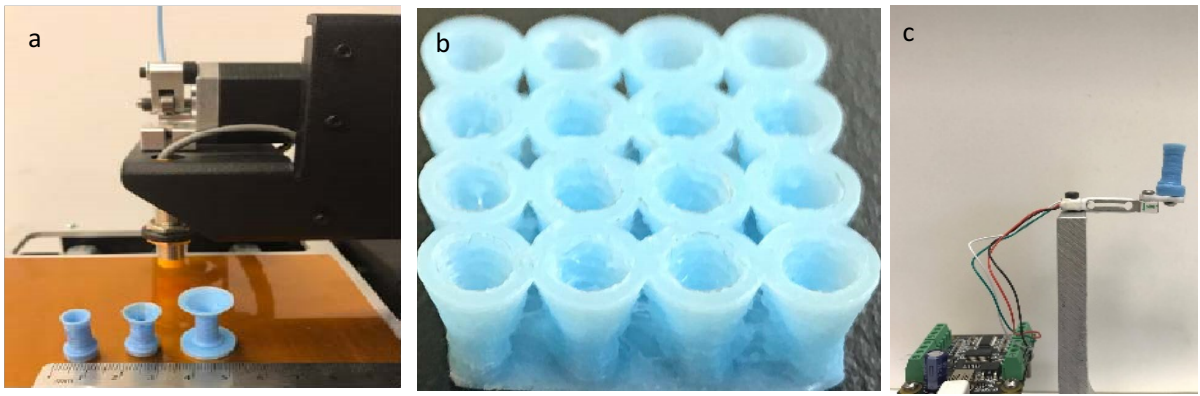

**Figure 4.** (a) Printrbot Simple Metal 3D Printer, printed microthruster nozzles with 3 different ratios of outer diameter to inner chamber diameter, (b) side view of a 4×4 thruster array, (c) measurement setup with Phidget force sensors.

In microthrusters, the nozzle inner diameter was 0.7 mm and outer diameter was 3 mm. The wall thickness was 1 mm. The microthrusters in array occupy 5×5 mm area. The microthruster chamber was cylindrical with 0.7 mm diameter and 2 mm height.

Figure 5, (a) shows the measured thrust data, and Figure 5, (b-h) demonstrates the ignition images extracted from high speed video recording. For the 2 mg mass of loaded Al-Bi(OH)$_3$ nanothermite generates a peak thrust of 0.31 N. The area under the curve represents the total impulse, summing up to $I_{total}$=0.7 mNs with specific impulse $I_{sp}$=37 s.

The total impulse can be tuned by changing the mass of thermite and geometry of microthruster chamber. The nanothermite mass can be reduced to 0.1 mg or increased up to 10 mg for single microthruster to ensure required thrust, $I_{total}$ and $I_{sp}$ for specific flight phases and modes. In case of Al-Cu(IO$_3$)$_2$ nanothermite, for 0.1 mg mass the total impulse was 0.2 mNs, which increased up to 20 mNs for the charge mass of 10 mg [51]. The specific impulse was 216 s for this formulation. The studied single microthruster and thruster arrays can be scaled to nano-, micro- and mini-satellites and aerial systems.

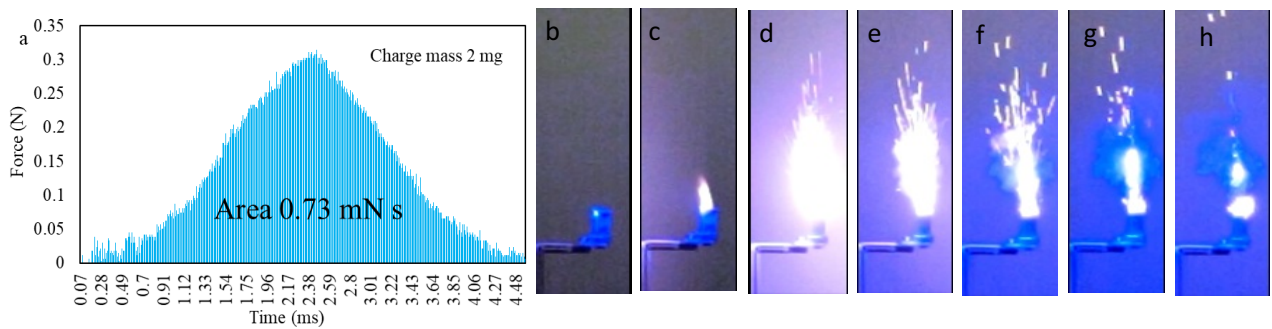

**Figure 5.** (a) Thrust by a single microthruster with 2 mg Al-Bi(OH)$_3$ nanothermite, (b-h) Consecutive frame snapshots of charge ignition extracted from high speed video recording at 980 fps.

*4.2 High power output actuators based on MWCNT/NGGs composite yarns*

The NGGs can release high-pressure gasses at microsecond timeframe, and the resulting rapid expansion of gaseous products can potentially generate a large-work, and a high-force actuation, if the NGGs are incorporated within the yarns made from Multi Walled Carbon Nanotube (MWCNT) sheets [52]. The MWCNTs possess excellent mechanical and physical properties, which are especially desired for applications where remarkable strength and high stiffness (reaching up to 300 MPa) of single-ply yarns can be utilized [53]. The MWCNT sheets extracted from forest can be scrolled to form yarns, which can incorporate various functional materials, where the composite yarns can act as actuators [54, 55]. This strategy was used in this work, to create nanocomposites made from twisted MWCNT yarns that incorporate NGGs based on iodine and bismuth oxidizers. From the side of a MWCNT forest, well aligned MWCNT sheets were produced by dry-spinning using a razor blade. The width of the sheet is proportional to the width of the synthesized MWCNT forest (Figure 6). For a typical yarn actuator, 2-cm-wide MWCNT sheets were placed between two rods separated by a distance of 7 cm. 15 MWCNT sheets were stacked on top of each other to increase the sheet strength during NGG coating. The weight of the stacked MWCNT sheets was ~0.8 mg.

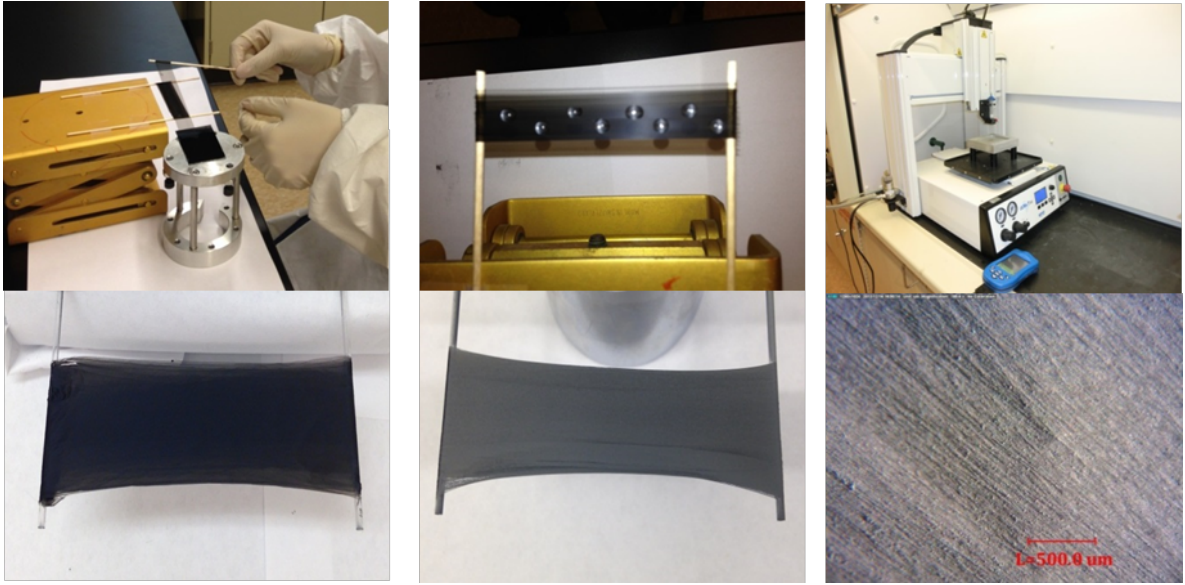

**Figure 6.** Preparation of MWCNT sheets, and coating with NGG using robot EFD-325TT equipped with a Paasche VL-SET airbrush system.

MWCNT sheets were coated with NGG materials by utilizing automated dispensing robot EFD-325TT equipped with a Paasche VL-SET airbrush system. The coating solution was prepared by suspending 25 mg/mL of NGG particles in isopropanol and sonicating for 30 minutes. The robot was programmed to coat the sheets in 30 second time intervals to allow the evaporation of isopropanol and control the MWCNT/NGG weight ratio (Figure 6). The laminar yarns were prepared by twisting the MWCNT/NGG sheets with a low speed motor under a constant speed 30 rpm. To form MWCNT/NGG yarns, the coated sheets were twisted into Archimedean type yarns, as described in [56], with 800-1000 turns/m per final yarn length (Figure 7). The yarns with coiled structure were produced by addition of further $2.5 \times 10^4$ turns/m [57].

The force stroke measurements were performed using Phidget force sensor calibrated for forces up to 1 N. The frequency of 50 Hz was used to monitor the force change over time. In this case, the yarn length was not allowed to change. An immobile base was used to attach one end of the yarn, while the other end was attached to the force sensor. Ignition of the yarns was initiated by creating 15 V potential difference across the yarn, which caused Joule heating and ignition of the yarn. In another set of experiments, the length of yarn was allowed to change, and the actuation force was measured with pre-calibrated cantilever.

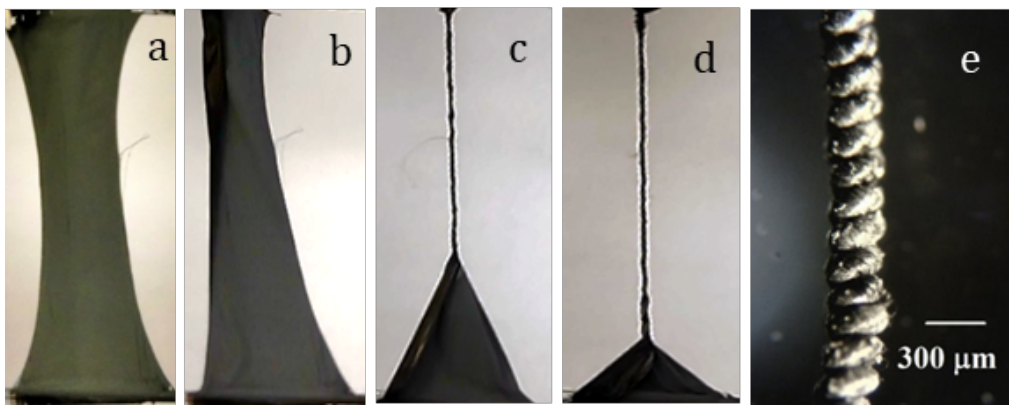

**Figure 7.** (a-d) The evolution of spin rotation of the NGG coated MWCNT sheets to form a yarn; and (e) Micrograph of the coiled yarn.

Upon ignition of the MWCNT/NGG twisted and coiled yarns, gaseous iodine rapidly escapes the yarn through spaces between MWCNT sheets, which increases the diameter of the yarn. This process is demonstrated using snapshots from high speed camera recording and SEM images in Figure 8, (a, b). During the timeframe of 8-34 ms, the diameter of the yarn increased by about 10 times, which caused a force stroke (~0.45 N) along the yarn length, Figure 8, (a), shown with blue arrows. The final diameter of yarn increased reaching 520 μm after cooling, which is a 90 % increase in comparison to 275 μm diameter value of yarn before the ignition.

The actuation stroke of the MWCNT/NGG twisted and coiled yarns was measured by attaching one end of the yarn to an immobile base, while the other end was attached to the force sensor. MWCNT/NGG twisted yarns which had mass ratio 1:2 generated a 0.45 N stroke force.

However, the heavily coiled yarns produced ~ 0.2 N force, Figure 8, (c). The twofold reduction of stroke force for coiled yarns compared to twisted yarns, can be due to the tight MWCNT structure, that does not allow the release of the iodine gas that leads to an increase in diameter (coiled yarns show less than 20 % diameter increase, as opposed to twisted yarns with over 90 % diameter increase after actuation). For both coiled and twisted yarns, the force impulse duration was between 0.05-0.15 s. After combustion, the coiled and twisted yarns have a remnant stress force of 0.08 N and 0.18 N, respectively, Figure 8, (c). Thus, after actuation, both coiled and twisted yarns produce stroke forces and generate a remnant stress along the yarn. When the NGGs contains excess amount of oxidizer, the yarns produce a higher stroke force of ~0.6 N, however, the excess oxidizer damages ~90 % of the MWCNT/NGG yarn integrity after ignition and actuation. In this case, no remnant stress is observed.

In order to evaluate the power of actuation for the MWCNT/NGG twisted and coiled yarns, we used a setup of lifting load of 2 g mass. While in the force stroke measurement the two ends of the yarn were fixed, in this setup during the actuation the bottom end of the yarn can rotate around the yarn axis. The measured specific power for coiled yarns was higher due to kinetic energy of torsional untwisting, which produced rotation of mass during uplift. During the yarn actuation, the specific power up to 4700 W/kg was reached. This value is 94 times higher than the power of a typical mammalian muscle (50 W/Kg [58]). The actuation of MWCNT/NGGs composite yarns work in ambient air, vacuum, and inert environment [52].

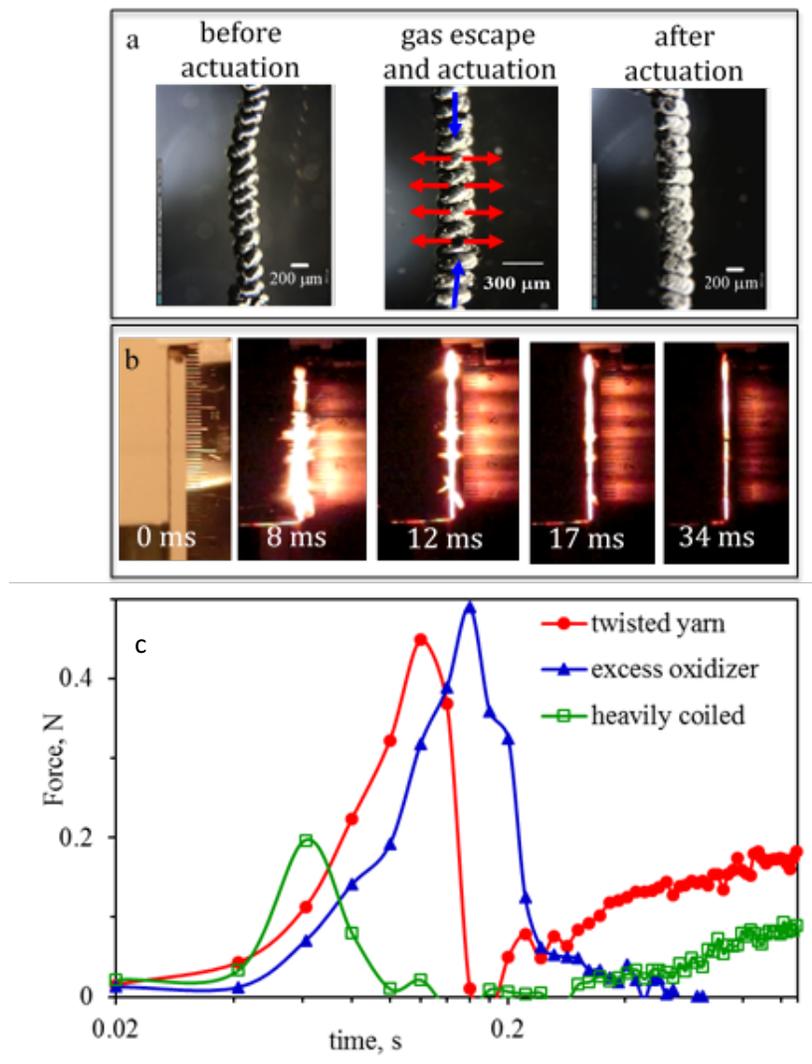

**Figure 8.** Yarn actuation: (a) The actuator yarn before, and after actuation, and the scheme of actuation (blue arrows) due to gas escape (red arrows), b) The snapshots extracted from 240 fps video recording (IR filtered), demonstrating the change of yarn diameter during gas generation, c) The force measurement over time for various yarns: simple twisted yarn, CNT/NGGs mass ratio 1:2, twisted yarn made with thermite containing 67 wt. % excess $I_2O_5$ oxidizer, and heavily coiled yarn, CNT/NGGs mass ratio 1:2.

The DSC-TGA analysis of the yarns after combustion/actuation at the heating rate of 20 º/min under 100 mL/min air flow was performed to quantitatively estimate the percentage of MWCNT burned, (Figure 9). Until 600 ºC only 4.7 wt. % mass reduction was observed. This mass can be contributed to iodine, some amount of which still remain in the yarn after actuation. The yarn starts to burn at around 600 ºC with 49 wt. % mass reduction and 5832 J/g energy release. Thus, in the yarn after actuation about half of the mass are MWCNT.

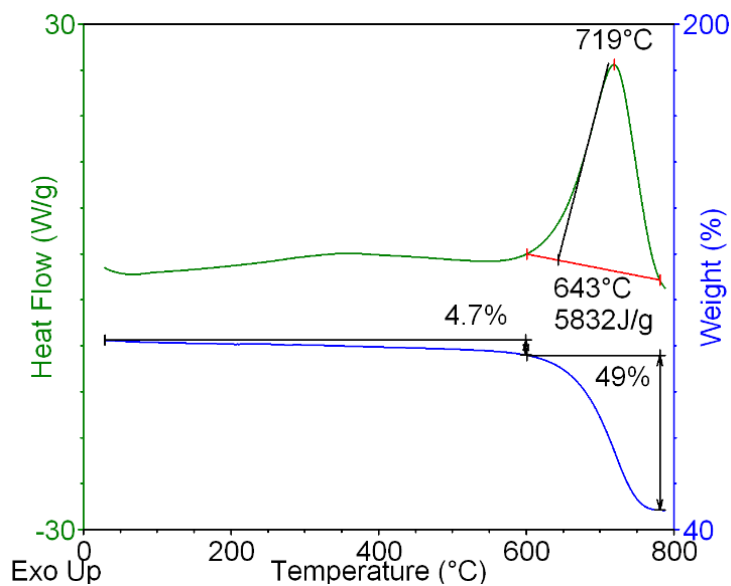

**Figure 9.** DSC-TGA analysis of yarn after actuation.

SEM images and EDS element mapping of the coiled and twisted MWCNT-NGG yarn show the morphology and distribution of the remaining aluminum and oxygen left after combustion along the yarns, (Figure 10, (a-e)). TEM image of the individual aluminum oxide nanoparticles reveal a size of ~ 50 nm and no interconnection with individual carbon nanotubes, Figure 10, (f).

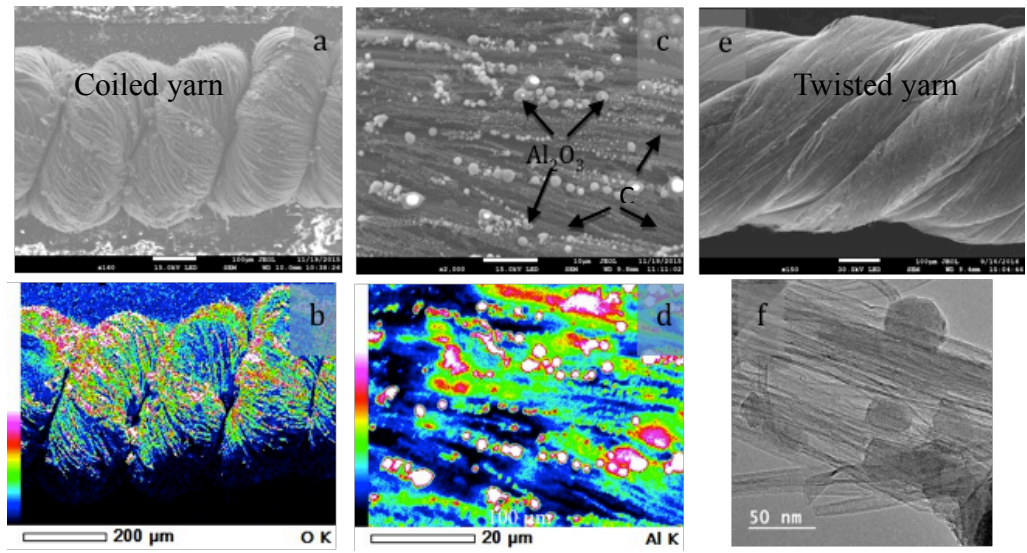

**Figure 10.** (a) SEM of a coiled yarn after actuation, (b) EDX mapping of oxygen atoms for area shown in (a); (c) SEM images at higher magnification for EDX distribution analysis; (d) distribution of Al atoms for area shown in (c); (e) SEM of 65 wt. % NGG containing twisted yarn after actuation; (f) Individual aluminum oxide nanoparticles around MWCNT.

*4.3 NGG based on iodine pentoxide oxidizer for biocidal agent defeat*

Traditionally, the iodine-based substances have been effective, simple and cost-effective for disinfection for hundreds of years. Iodine has been used in France during World War I to disinfect water; US Army used Globalin (tetraglycine hydroperiodide) tablets during World War II. The disinfection based on iodine has been used by NASA in space flights [59]. Since early 50s extensive research has been conducted exploring the efficiency of disinfection for various forms of iodine [60-62]. Iodine is rapidly bactericidal, fungicidal, tuberculocidal, virucidal, and sporicidal [63].

For testing the biocidal properties of nanothermites, plexiglass reaction chamber with a volume of 49.26 L and a wall thickness of 1.2 cm with a weight of 3 kg was used. The samples were placed at various distances and orientation demonstrated in Figure 11, (a). The charge was placed at 8 cm height and was electrically ignited using a variable autotransformer from Staco Energy Produces. An Escherichia coli (E. coli) bacteria strain HB101 K-12 was used, which is not pathogenic like the strain O157 H7, which has been implicated in some occasions in food poisoning. HB101 K-12 strain has been genetically modified so it can only be grown in enriched medium. We exposed them to nano-thermite disinfection after 1 hours of incubation of bacteria, placing the agar plates coated with bacteria at various orientations in the chamber shown in Figure 11, (a). The agar plates are placed in an incubator for 24 hours at 37 ºC after the biocidal treatment. If the treatment is successful there will not be any visible growth on the agar plate. The *E.coli* colonies are counted and the number of visible colonies is compared with two control agar plates (Figure 11, (b)) that were made but were not exposed to the biocidal gas and were allowed to grow. Thus, the biocidal effect of $I_2O_5$-Al NGG mixture weight, exposure time, as well as sample distance and orientation from charge was investigated. According to calculation, 0.1g mixture generates 0.53 g atomic iodine [32]. All samples were completely disinfected, regardless orientation or distance from the charge (Figure 11, (c), compare to Figure 11, (b)). Calculations show that minimal effective concentration of iodine over the infected area is 22 mg/m$^2$ (Figure 11, (d)).

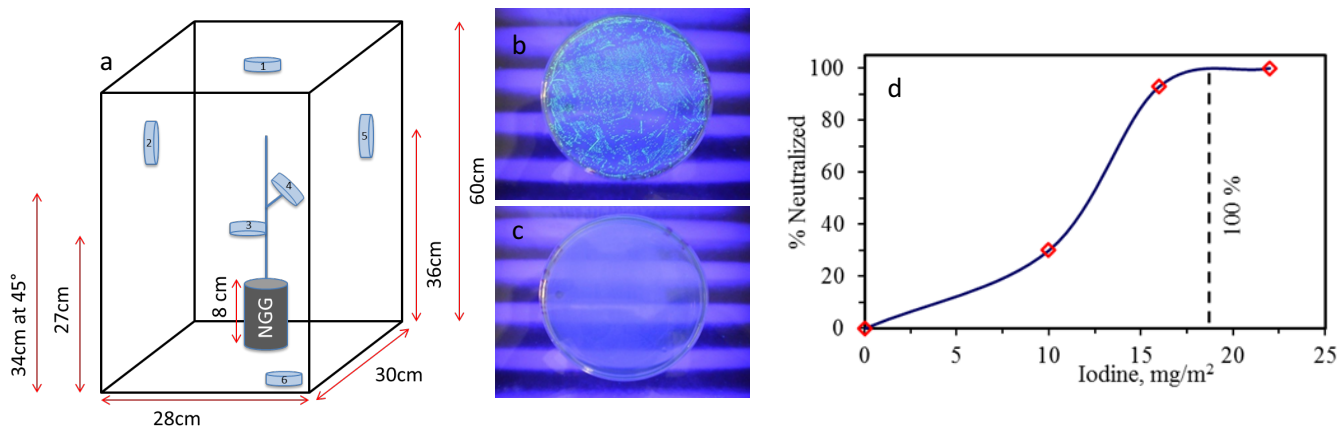

**Figure 11.** (a) The experimental setup for biocidal performance; (b) Control sample; (c) Sample exposed to nanothermite disinfection; (d) Statistical curve of killing abilities of active iodine generated by nanothermite reaction

## 5. Summary

This chapter reported the recent research findings of nanoenergetic gas generator systems containing bismuth and iodine oxidizers, and tuning the reactivity of the systems by altering the oxidizer particle size, as well as shape. The iodine pentoxide, bismuth hydroxide and bismuth trioxide surpass other traditionally known oxidizers in thermite compositions in terms of providing superior pressure discharge values up to 14.8 kPam$^3$/g. These NGG formulations can be used in new emerging technological applications such as microthrusters, providing milli-Newton-second thrust resolution, as well as linear composite actuators with high power output up to 4700 W/kg, and biocidal agents with excellent bactericidal properties, where the minimal effective concentration over the infected area was 22 mg/m$^2$.